\documentclass[12pt]{article}
\usepackage{epsfig}
\usepackage{amsfonts}
\usepackage{amssymb}
\title{Unitarity in the noncommutative theories. }
\author{Piotr Kosi\'nski\thanks{supported by the grant no. 1 P03B 021 28 of Polish Minister of Science  } , 
 Pawe{\l} Ma\'slanka$^*$ \\
Department of Theoretical Physics II \\
University of {\L}\'od\'z \\
Pomorska 149/153, 90 - 236 {\L}\'od\'z/Poland.}
\date{}

\begin{document}
\maketitle
\begin{abstract}
Simple argument in favour of unitarity, to all orders, of space-like noncommutative field theory is given.
\end{abstract}

\newpage
The noncommutative field theories, being non-local, lead to a number of new phenomena such as UV/IR mixing or acausal
behaviour \cite{b1} $\div $\ \cite{b4}. An important question is also whether such theories are unitary. The common 
belief is that the answer to the latter is positive for space-like noncommutative theories ( i. e. such that 
$\theta^{i0}=0$)\ while time-like noncommutativity ($\theta ^{i0}\not=0)$\ leads to nonunitarity. The arguments 
supporting such conclusion can be given within the field theory framework. Namely, the theories with space-like 
noncommutativity are non-local in space but are local in time. Therefore, the Hamiltonian formalism can be constructed
which, in principle, gives rise to unitary time evolution. On the contrary, the theories with time-like noncommutativity, 
containing time derivatives of arbitrary order, are non-local in time which may lead to breakdown of unitarity.
This may happen even within the perturbational regime where some modes are exluded from the very begining.

The field theory arguments can be supported by those coming from string theory. It is known,\cite{b5} $\div $\ \cite{b7},
that noncommutative field theory describe the low energy excitation of a D-brane in the presence of a background magnetic 
field. In this limit the massive open strings and the closed strings decouple and one obtains the consistent reduction
of the full string theory to field theory with space-like noncommutativity; since the former is unitary the latter should 
be unitary as well. On the other hand the field theory with time-like noncommutativity can not be obtained as a consitent 
limit of string theory \cite{b8} $\div $\ \cite{b10}.The physical interpretation of the breakdown of unitarity in
 such theories in terms of production of tachionic particles has been proposed in \cite{b26}. 

The above general arguments can be supported by actual perturbative calculations,\cite{b11} $\div $\ \cite{b16}, 
mainly on one loop level.

Let us also note that the theories with light-like noncommutativity, which lie on the borderline between time-like 
and space-like noncommutative theories, were proven to be unitary \cite{b27}.

\vspace{8pt} 

In the present note we show how the unitarity of space-like noncommutative field theories can be proven in the 
simple manner to any order of  perturbation theory. To this end we use the elegant approach developed by
 Veltman \cite{b17} and 't Hooft and Veltman \cite{b18}. It is based on configuration space version of Feynman rules.
 To any Feynman diagram with n-vertices the appropriate function $F(x_1,...,x_n)$\ is ascribed in the way following 
from the relevant Feynman rules.

 Being the product of singular functions ( and integrated further over $x's$) $F$\ has
 to be regularized somehow.
 In what follows some regularization procedure is assumed which respects the decomposition (\ref{w1}) below. One can,
 for example, integrate first over energy variables and impose some cut-off on remaining momentum integrations or use
 dimensional regularization for Wick-rotated integrals. It should be stressed here that, once the proper ( in the sense
 described above ) regularization is imposed, the method sketched below gives precise and nonformal proof that the 
regularized amplitude has the analytic structure respecting unitarity ( Cutkosky rules), contrary to the formal argument
based on the existence of hamiltonian formalism. Moreover, along the same lines, adopting further arguments of 't Hooft 
and Veltman to noncommutative situation, one shows that the possible divergencies must be local in time, so they cannot
 correspond to the new propagating degrees of freedom. An unpleasant aspect of noncommutative field theories is related to 
 renormalizability which, due to UV/IR mixing, becomes questionable ( we comment on it below ). However, the very problem 
of unitarity appears already on the level of regularized theories. This is well known also in the standard case. Given even
a nonrenormalizable lagrangian one can ask whether it gives rise to unitary theory. The answer is, roughly speaking, that 
unitarity is preserved provided the lagrangian is hermitian. This statement has the precise meaning because the divergencies, 
proliferating with growing order of perturbation theory, have trivial analytical structure so they do not influence unitarity
 but only make the predictive power of the theory doubtful due to the infinite number of parameters. This reasoning provides, 
for example, the basis for effective lagrangian method which allows to produce unitary, causal ond covariant amplitudes 
defined up to some momentum polynomials \cite{b30}. In the noncommutative case we are faced with a similar situation. It can happen that the
structure of divergencies gets more and more complicated with growing order of perturbation expansion and cannot be tamed 
by nonlocal counterterms of the same form as the terms appearing in the lagrangian ( for some theories this happens already 
on the two-loops level ). However, what the unitarity proof does show is that the necessary counterterms do not violate the 
analyticity structure imposed by unitarity condition which takes into account only those degree of freedom which appear
 in the free part of lagrangian. 

As mentioned above in the noncommutative case the renormalizability becomes, due to UV/IR mixing, 
a subtle and generally speaking, 
unsolved problem.  There are only some partial results. It is known that $\phi ^4$-theory is renormalizable to two-loops
 order \cite{b19},\cite{b31}. As far as complex-field $\phi ^4$-theory is considered, it is known to be renormalizable
 on one-loop level for the special choice of coupling constants \cite{b32},\cite{b33}; also scalar electrodynamics is 
renormalizable on one-loop level provided the scalar potential has a specific form (\cite{b32},\cite{b34}). Further, the
 list of one-loop renormalizable cases includes N=2 SUSY Yang-Mills theory \cite{b32},\cite{b34}.

It is much more difficult to say something about renormalizability to any order of perturbation theory. There are strong 
indications, based on generalized power-counting theorem (\cite{b21},\cite{b22}) that the noncommutative Wess-Zumino
model is  renormalizable (see also \cite{b23},\cite{b24}); the same seems to hold true for special form of complex 
$\phi ^4$-theory \cite{b21},\cite{b22}. What concerns the real $\phi ^4$-theory the Wilson renormalization group-inspired 
analysis \cite{b20} indicates that the theory is renormalizable provided some partial resummation procedure is applied.

 \vspace{8pt}

$F(x_1,...,x_n)$ defined above, when multiplied by the wave functions of incoming and outgoing particles and integrated 
over $x_1,...,x_n$, contributes to the relevant matrix element. Moreover, to account for $S^+$\ marix elements
 the more general functions are defined with some of the variables $x_1,...,x_n$\ underlined. These new functions
are obtained from $F(x_1,...,x_n)$\ by replacing $\Delta _F(x_k-x_i)$\ by $\Delta ^+(x_k-x_i)$, $\Delta ^-(x_k-x_i)$\
or $\bar \Delta _F(x_k-x_i)$\ if, respectively, $x_k$\ but not $x_i$\ is underlined , $x_i$\ but not $x_k$\ is 
underlined or both $x_k$\ and $x_i$\ are underlined; further one replaces the notorious factor $i$\ by $-i$\ for any
underlined vertex $x_k$.

Taking into account the decomposition
\begin{eqnarray}
\Delta _F(x)=\Theta (x^0)\Delta ^+(x)+\Theta (-x^0)\Delta ^-(x) \label{w1}
\end{eqnarray}
the following relation is easily shown to hold
\begin{eqnarray}
\sum\limits_{underlinings} F(x_1,...,\underline{x}_i,...,\underline{x}_j,...,x_n)=0 \label{w2}
\end{eqnarray}
where the summation goes over all possible ways the variables are underlined. \\
Again, eq.(2), when multiplied by the approprate wave functions and integrated over $x_1,...,x_n$, represents, on the level 
of individual diagrams, the unitarity relation $T-T^+=iT^+T$\ provided: 

{\bf i)} the modifield (i.e. those corresponding to underlined vertices) Feynman 

rules describe the $S^+$\ matrix elements\\ 
and 

{\bf ii)} the $\Delta ^+$\ functions are equal to the sums over intermediate states.

\vspace{8pt}

 Given a particular theory both conditions 
can be easily checked (for example, (i) is satisfied provided the Lagrangian is hermitian). The only difficulty is 
encountered if we are dealing with gauge theories where the sum over intermediate states includes nonphysical states; then 
one has to use Ward identities to show that the contributions from nonphysical states do cancel. The key point of the 't 
Hooft - Veltman approach is the following. Eq.(2) is the immediate consequence of the so called largest-time equation which, 
in turn, follows by considering any particular time-ordering of $x_1,...,x_n$\ and taking into account eq.(1) as well as the 
extra minus sign related to any underlined vertex. The largest-time equation and, consequently, eq.(2) are the identities 
involving products of propagators. However, the derivation of unitarity condition is valid for theories with derivative 
couplings of finite orders because a finite order differential operator is local and can be applied to the largest-time
equation even if the former contains time 
derivatives. The noncommutative theories contain, in their couplings, the derivatives of arbitrary orders,
i.e. nonlocal operators. This is why the largest-time 
equation may be violated when such couplings enter the theory. There is, however, one exception. The space components 
of $x's$\ play no role in the largest-time equation. Therefore, one can apply to both sides of the latter any operator, 
local or not, provided it is local in time variables. Taking into account that the quadratic part of the action of noncommutative 
field theory coincides with its commutative counterpart (i.e. the propagators and wave functions of both are the same) while 
the interaction Lagrangian is hermitian we conclude that the noncommutative field theory with $\theta ^{0i}=0$\ is unitary.

For illustration we consider the scalar $\phi ^4$-theory. For $\phi ^4$-theory the configuration-space 
Feynman rules can be summarized as follows: the topology of graphs as well as all symmetry factors are the same as for its 
commutative counterpart. Given any graph the "commutative" amplitude is built as a product of Feynman propagators corresponding 
to all internal lines of the graph and the wave functions of incoming and outgoing particles for all external lines: moreover, 
there is an additional $-i\lambda $\ factor for each vertex. The modifications necessary to take into account the 
noncommutativity of the space-time can be summarized as follows: for any internal or external line $l$\ one introduces an 
additional parameter $\xi ^{\mu }_l$\ and makes the replacements $\Delta _F(x_i-x_j)\longrightarrow \Delta _F(x_i-x_j+
\xi _l)$\ for an internal line $l$\ and $exp(\pm ip_lx_i) \longrightarrow exp(\pm ip_l(x_i+\xi _l))$ for an external line $l$. Then one constructs the 
configuration space amplitude according to the rules of commutative theory using, however, the modifield propagators and 
wave functions. The final step consists in applying to the resulting expression the product of the differential operators, 
one for each vertex, of the following form:
\begin{eqnarray} 
&& O_i=\frac{1}{3}(\cos(\frac{1}{2}\theta ^{\mu \nu }\frac{\partial }{\partial \xi ^{\mu }_{l_1}}\frac{\partial }{\partial \xi ^{\nu }
_{l_2}})\cos(\frac{1}{2}\theta ^{\mu \nu }\frac{\partial }{\partial \xi ^{\mu }_{l_3}}\frac{\partial }{\partial \xi ^{\nu }_{l_4}})+ \\ \nonumber
&& +\cos(\frac{1}{2}\theta ^{\mu \nu }\frac{\partial}{\partial \xi ^{\mu }_{l_1}}\frac{\partial }{\partial \xi ^{\nu }_{l_3}})\cos(\frac{1}{2}\theta ^{\mu \nu }\frac{\partial }
{\partial \xi ^{\mu }_{l_2}}\frac{\partial }{\partial \xi ^{\nu }_{l_4}})+ \\ \label{w3}
&& +\cos(\frac{1}{2}\theta ^{\mu \nu }\frac{\partial }{\partial \xi ^{\mu }_{l_1}}\frac{\partial }{\partial \xi ^{\nu }_{l_4}})
\cos(\frac{1}{2}\theta ^{\mu \nu }\frac{\partial }{\partial \xi ^{\mu }_{l_2}}\frac{\partial }{\partial \xi ^{\nu }_{l_3}})) \nonumber
\end{eqnarray}
At the very end one puts $\xi _l=0$. \\
Assume now that we are dealing with the space-like noncommutative theory, i.e. $\theta ^{i0}=0$; then from the very begining 
one can put $\xi ^0_l=0$. We see that the largest-time equation, which is based on the decomposition (1), holds for "
commutative" amplitude with modifield propagators and wave functions.  Consequently eq.(2) does hold identically in $\xi ^k_l$. 
Taking into account that the operators $O_i$, being real, correspond to both underlined and notunderlined vertices one 
applies the products of them to both sides of modified eq.(2). Putting $\xi _l=0$\ and integrating over all $x's$\ one 
arrives at the unitarity relation for noncommutative $\phi ^4$-theory. \\
Similar proof can be given for other space-like noncommutative theories. Let us take the noncommutative Q.E.D.
 Apart from the 
obvious algebraic complications coming from the fact that it describes the charged particles with spin, the new feature 
emerges: due to the existence of three - and four - photon vertices the topology of Feynman graphs is much more complicated 
than in the commutative case. However, once the revelant graphs are classified one can follow the same line of reasoning 
as above. Obviously, as we have mentioned above, the relations derived in this way describe unitarity in the whole Hilbert 
space of states. In order to prove unitarity in  the physical subspace one would have to use the relevant Ward identities.

\vspace {16pt}

{\Large\bf Aknowledgment}

We thank R. Zwicky and J. Gomis for bringing  to our attention Refs. \cite{b26} and \cite{b27}, respectively.


\begin{thebibliography}{99}
\bibitem{b1}
S. Minwalla, M.V. Raamsdonk, N. Seiberg, JHEP {\bf 0002} (2000), 020
\bibitem{b2}
M.V. Raamsdonk, N. Seiberg, JHEP {\bf 0003} (2000), 035
\bibitem{b3}
A. Matusis, L. Susskind, N. Toumbas, JHEP {\bf 0012} (2000), 002
\bibitem{b4}
N. Seiberg, L. Susskind, N. Toumbas, JHEP {\bf 0006} (2000), 044
\bibitem{b5}
A. Connes, M.R. Douglas, A. Schwartz, JHEP {\bf 9802} (1998), 003
\bibitem{b6}
M. R. Douglas, C. Hull, , JHEP {\bf 9802} (1998), 008
\bibitem{b7}
N. Seiberg, E. Witten, JHEP {\bf 9909} (1999), 032
\bibitem{b8}
N. Seiberg, L. Susskind, N. Toumbas, JHEP {\bf 0006} (2000), 021
\bibitem{b9}
R. Gopakumar, J. Maldacena, S. Minwalla, A. Strominger, JHEP {\bf 0006} (2000), 036
\bibitem{b10}
J.L.F. Barbon, E. Rabinovici, Phys. Lett. {\bf B486} (2000), 202
\bibitem{b26}
L. Alvarez-Gaume, J.L.F. Barbon, R. Zwicky, JHEP {\bf 0105} (2001), 057
\bibitem{b11}
J. Gomis, T. Mehen, Nucl. Phys.{\bf B591} (2000), 265
\bibitem{b12}
T. Mateos, A. Moreno, Phys.Rev. {\bf D64} (2001), 047703
\bibitem{b13}
A. Bassetto, L. Griguolo, G. Nardelli, F. Vian, JHEP {\bf 0107} (2001), 008
\bibitem{b14}
C-S. Chu, J. Lukierski, W. Zakrzewski, Nucl. Phys. {\bf B632} (2002),219
\bibitem{b15}
K. Morita, Y. Okumura, E. Umezawa, Prog. Theor. Phys. {\bf 110} (2003), 989
\bibitem{b16}
A. Smailagic, E. Spalluci, J. Phys. {\bf A37} (2004), 1
\bibitem{b27}
O. Aharony, J. Gomis, T. Mehen, JHEP{\bf 0009} (2000), 023 
\bibitem{b17}
M. Veltman, Physica {\bf 29} (1963), 186
\bibitem{b18}
G.'t Hooft, M. Veltman, "Diagrammar", CERN yellow report 73-9, 1973
\bibitem{b30}
S. Weinberg, Physica {\bf 96A} (1979), 327\
\bibitem{b19}
I. Ya. Aref'eva, D. M. Belov, A. S. Koshelev, Phys. Lett. {\bf B476} (2000), 431
\bibitem{b31}
A. Micu, M.M. Sheikh-Jabbari, JHEP{\bf 0101} (2001), 025
\bibitem{b32}
 I.Ya. Aref'eva, D.M. Belov, A.S. Koshelev, O.A. Rytchkov,  Phys. Lett. {\bf B487} (2000), 357
\bibitem{b33}
 I.Ya. Aref'eva, D.M. Belov, A.S. Koshelev,"A Note on UV/IR for Noncommutative Complex Scalar Field", hep-th/0001215
 \bibitem{b34}
 I.Ya. Aref'eva, D.M. Belov, A.S. Koshelev, O.A. Rytchkov, Nucl. Phys. Proc. Suppl. {\bf 102} (2001), 11
 \bibitem{b21}
I. Chepelev, R. Roiban, JHEP {\bf 0005} (2000), 037
\bibitem{b22}
I. Chepelev, R. Roiban, JHEP {\bf 0103} (2001), 001
\bibitem{b23}
H. O. Girotti, M. Gomes, V.O. Rivelles, A. J. da Silva, Nucl. Phys. {\bf B587} (2000), 299
\bibitem{b24}
I. L. Buchbinder, M. Gomes, A. Yu. Petrov, V.O. Rivelles, Phys. Lett. {\bf B517} (2000), 191
\bibitem{b20}
L. Griguolo, M. Pietroni, JHEP {\bf 0105} (2001), 032
 
\end{thebibliography}
\end{document}